\documentclass[aip,jap,preprint,graphicx]{revtex4-1}
\usepackage{graphicx}% Include figure files
\usepackage{dcolumn}% Align table columns on decimal point
\usepackage{bm}% bold math

\usepackage{amsmath}
\usepackage{amssymb}

\begin{document}

%\preprint{}

\title{Definition of the interlayer interaction type in magnetic multilayers analyzing the shape of the ferromagnetic resonance peaks} %Title of paper


\author{O.G.~Udalov}
\email[]{oleg.udalov@csun.edu}
%\thanks{}
\affiliation{California State University, Northridge, CA, USA}
\affiliation{Institute for Physics of Microstructures RAS, Nizhny Novgorod, Russia}

\author{A.A.~Fraerman}

%\email[oleg.udalov@csun.edu]{Your e-mail address}
%\thanks{}
\affiliation{Institute for Physics of Microstructures RAS, Nizhny Novgorod, Russia}

\author{E.S.~Demidov}
%\thanks{}

\affiliation{Lobachevsky State University of Nizhni Novgorod, Nizhny Novgorod, Russia}


\date{\today}

\begin{abstract}
We present theoretical study of ferromagnetic resonance in a system of two coupled magnetic layers. We show that an interaction between the layers leads to the occurrence of the so-called Fano resonance. The Fano resonance changes the shape of the ferromagnetic resonance peak. It introduces a peak asymmetry. The asymmetry type is defined by the sign of the interaction between the magnetic layers. Therefore, studying the shape of the ferromagnetic resonance peaks one can define the type of the interlayer coupling (ferromagnetic or antiferromagnetic). We show that using numerical simulations one can estimate a magnitude of the interaction by fitting the asymmetric resonance peaks.
\end{abstract}


\maketitle %\maketitle must follow title, authors, abstract and \pacs

\section{Introduction}\label{Sec:Intro}

Ferromagnetic resonance (FMR) is a powerful tool for studying of magnetic multilayer structures~[\onlinecite{Baberschke2003,Passamani2006,Lesnik2006,Lesnik2007,PhysRevApplied.8.044006, doi:10.1021/acsami.7b00284,doi:10.1021/acsnano.7b01547,0953-8984-15-4-204,doi:10.1063/1.3176901,PhysRevB.84.184438}]. The FRM method allows to obtain the information on the magnetization magnitude and magnetic anisotropy of each layer. It can be also used for studying of the interlayer coupling. A lot of efforts were spent on investigation of the coupling in the systems with magnetic layers separated by a metallic non-magnetic spacer~[\onlinecite{Baberschke2003,0953-8984-15-43-003,doi:10.1063/1.3033519,0953-8984-15-4-204,0953-8984-15-5-301,Saitovitch2011,Kaiser2017}]. In this case the interlayer coupling is strong enough. This makes it relatively easy to define the coupling sign and magnitude studying shifts of FMR peaks. 

The situation is different for magnetic multilayers where ferromagnetic films are separated by an insulating spacer leading to a much weaker interlayer coupling~[\onlinecite{Kawakami2010,Dieny2011}]. Measuring the coupling in this case is a tricky issue. The mutual shift of FMR peaks corresponding to different layers is small comparing to the peaks width~[\onlinecite{Lesnik2006,Lesnik2007}]. The situation becomes even more complicated when resonant fields (frequencies) of the peaks are close to each other. In this case a completely different approach is needed.

In the present work we propose to define the interlayer interaction sign and magnitude by studying the FMR peaks shape rather than the shift. We will show that the interaction induces an FMR peaks asymmetry. Such an asymmetry can be considered as the Fano resonance~[\onlinecite{Satanin2006}] in a magnetic multilayer. Studying the shape of this asymmetry one can define the interaction sign and magnitude. Such a method is particularly useful when resonance frequencies of two interacting layers are close to each other. 

Studying of the interaction sign and magnitude with the conventional method based on the FMR peaks shift requires a reference sample without the interlayer interaction. This allows to measure the peak shift. The approach based on the peak shape does not have such a disadvantage. One can define the interaction sign and magnitude using a single sample. 

The paper is organized as follows. In the Sec.~\ref{Sec:SimpleModel} we analyse a simplified model in which two magnetic moments are placed into a strong magnetic field. Such a model allows analytical consideration providing the insight into the physics behind the FMR peak shape (asymmetry). In Sec.~\ref{Sec:NumSim} we study numerically magnetic bilayer system (NiFe/Co)  with an arbitrary orientation of the external magnetic field.

\section{Simplified model}\label{Sec:SimpleModel}
\begin{figure}
	\includegraphics[width=0.5\columnwidth]{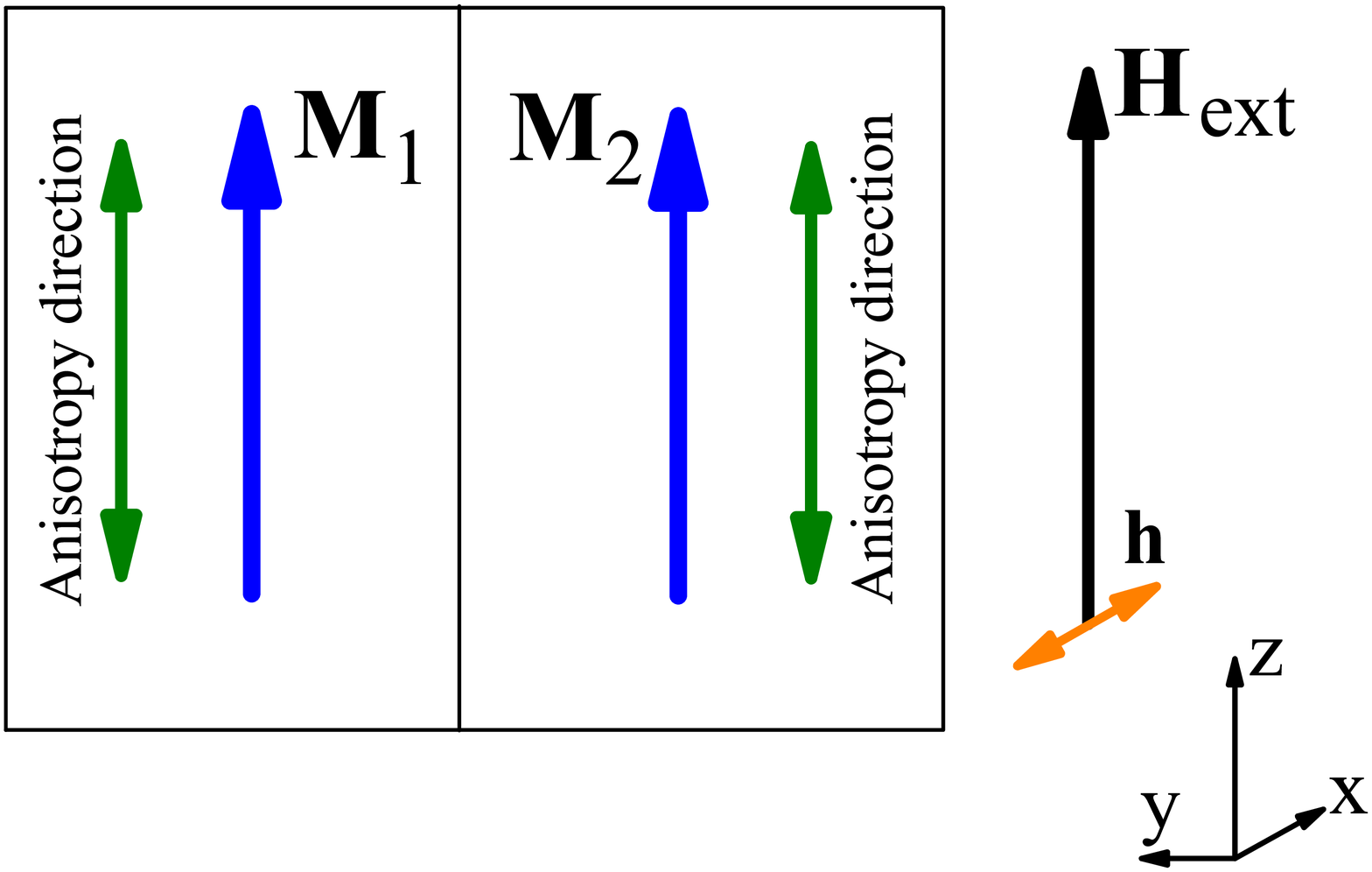}
	\caption{A model system. Two magnetic moments placed in an external magnetic field $\mathbf H_{\mathrm{ext}}$. An alternating magnetic field $\mathbf h$ is applied perpendicular to $\mathbf H_\mathrm{ext}$. $\mathbf M_{1,2}$ show equilibrium orientation of the magnetic moments. \label{Fig:SysModel}}%
\end{figure}

In this section we consider a simplified model of two coupled magnetic moments. We calculate dissipation (FMR signal) in this system and demonstrate how the asymmetric peak of absorption appears. Consider two ferromagnetic (FM) films with uniform magnetizations $\mathbf M_{1,2}$ (see Fig.~\ref{Fig:SysModel}). For simplicity we assume that the magnetic moments of both layers are the same $|\mathbf M_{1,2}|=M_0$. There is a uniaxial anisotropy in each film along the z-axis.  It can be induced by a demagnetizing field or by an internal anisotropy. The anisotropy constants are $\lambda_{1,2}$. An external magnetic field $\mathbf H_\mathrm{ext}=H_0\mathbf z_0$ is applied to the system. There is also a weak high-frequency alternating field along the x-axis $\mathbf h(t)=h(t)\mathbf x_0$. Magnetic films interact with each other. The interaction energy is given by the expression
\begin{equation}\label{Eq:InterGen}
E_\mathrm{int}=-\tilde J(\mathbf M_{1}\mathbf M_{2}).
\end{equation}

We linearize the Landau-Lifshitz-Gilbert (LLG) equations for both magnetic moments $\mathbf M_{1,2}$ in the vicinity of equilibrium positions $\mathbf M_{1,2}=M_0 \mathbf z_0$. The equations take the form
\begin{equation}\label{Eq:LLGGen}
\left\{\begin{split}
&\dot m_{1x}=-H_1 m_{1y}-J(m_{1y}-m_{2y})-\tilde\alpha_1 \dot m_{1y},\\
&\dot m_{1y}=H_1 m_{1x}-J(m_{2x}-m_{1x})+\tilde\alpha_1 \dot m_{1x}-h,\\
&\dot m_{2x}=-H_2 m_{2y}+J(m_{1y}-m_{2y})-\tilde\alpha_2 \dot m_{2y},\\
&\dot m_{2y}=H_2 m_{2x}+J(m_{2x}- m_{1x})+\tilde\alpha_2 \dot m_{2x}-h.\\
\end{split}\right.
\end{equation}
Here $\mathbf m_{1,2}$ are the corrections to the equilibrium magnetizations normalized by $M_0$, the magnitude of the effective field acting on the layers are $H_{1,2}=\gamma(H_0+2\lambda_{1,2}M_0)$, $J=\gamma \tilde J$ is the interaction constant multiplied by the gyromagnetic ratio $\gamma$. The renormalized damping constants are $\tilde\alpha_{1,2}$. The system Eq.~(\ref{Eq:LLGGen}) can be transformed into two second order equations of the form 
\begin{equation}\label{Eq:LLGGen1}
\left\{\begin{split}
&\ddot m_{1x}+\alpha_1\dot m_{1x}+\omega_1^2 m_{1x}=A_1m_{2x}+D_1\dot m_{2x}+h_1,\\
&\ddot m_{2x}+\alpha_2\dot m_{2x}+\omega_2^2 m_{2x}=A_2m_{1x}+D_2\dot m_{1x}+h_2,\\
\end{split}\right.
\end{equation}
where we introduced the following notations
\begin{equation}\label{Eq:Notations}
\begin{split}
&\alpha_{1,2}=\frac{2\tilde\alpha_{1,2}(H_{1,2}+J)}{1+\tilde\alpha_{1,2}^2}\approx 2\tilde\alpha_{1,2}(H_{1,2}+J),\\
&\omega_{1,2}^2=\frac{(H_{1,2}+J)^2+J^2}{1+\tilde\alpha_{1,2}^2}\approx H_{1,2}^2+2J H_{1,2},\\
&A_{1,2}=\frac{(H_{1}+H_2)J+2J^2}{1+\tilde\alpha_{1,2}^2}\approx (H_{1}+H_2)J,\\
&D_{1,2}=\frac{J(\tilde\alpha_1+\tilde\alpha_2)}{1+\tilde\alpha_{1,2}^2}\approx 0,\\
&h_{1,2}=\frac{H_{1,2}h+\tilde\alpha_{1,2}\dot h}{1+\tilde\alpha_{1,2}^2}\approx H_{1,2}h.\\
\end{split}
\end{equation}
Equations~(\ref{Eq:LLGGen1}) describe the system of two coupled oscillators with the resonant frequencies $\omega_{1,2}$. There are two types of coupling between the oscillators. We assume that the damping is weak ($\tilde\alpha_{1,2}\ll 1$) which is often the case for ferromagnets. In this limit one can neglect the dissipative coupling terms $D_{1,2}\dot m_{1,2x}$. Also the retarded external excitation $\tilde \alpha_{1,2}\dot h$ can be omitted. For our purposes we can also neglect $\tilde\alpha_{1,2}^2$ in denominators in Eqs.~(\ref{Eq:Notations}). We assume that the coupling between the films $J$ is weak comparing to the effective fields $H_{1,2}$. Therefore, we keep only the terms linear in $J$. 

A response of the system to a periodic external field $h_{1,2}=h^{(0)}_{1,2}e^{i\omega t}$ can be represented as $m_{1,2x}(t)=m_{1,2}e^{i\omega t}$. The complex amplitudes  $m_{1,2}$ are given by
\begin{equation}\label{Eq:Amplitudes}
\left\{\begin{split}
&m_1=\frac{(\omega_2^2-\omega^2+i\alpha_2\omega)h_1^{(0)}+A_1 h_2^{(0)}}{(\omega_2^2-\omega^2+i\alpha_2\omega)(\omega_1^2-\omega^2+i\alpha_1\omega)-A_1A_2},\\
&m_2=\frac{(\omega_1^2-\omega^2+i\alpha_1\omega)h_2^{(0)}+A_2 h_1^{(0)}}{(\omega_2^2-\omega^2+i\alpha_2\omega)(\omega_1^2-\omega^2+i\alpha_1\omega)-A_1A_2}.\\
\end{split}\right.
\end{equation}

\subsection{Layers with essentially different damping, but the same resonant frequencies} 

Lets now further simplify our consideration assuming that $\alpha_2=0$ and $\omega_1=\omega_2$. This means that $H_{1}=H_2$, $h_{1}^{(0)}=h^{(0)}_2$, and $A_1=A_2=A$. Next we assume that the interaction is weak comparing to the damping ($\alpha_1\gg A/\omega_1$). In this case the oscillation amplitude of the first layer magnetization is given by 

\begin{equation}\label{Eq:Amplitude1}
\begin{split}
&|m_1|^2=\frac{((\omega_2^2-\omega^2)+A)^2 (h_2^{(0)})^2}{(\omega_2^2-\omega^2+A)^2(\omega_2^2-\omega^2-A)^2+\omega^2\alpha_1^2(\omega_2^2-\omega^2)^2}.
\end{split}
\end{equation}

In the case of no interaction ($A=0$) we have an ordinarily resonance peak with the frequency $\omega_2-\alpha_1^2/(4\omega_2)$. Introduction of the finite interaction $A$ leads to additional shift of the peak, but we can neglect it when $\alpha_1\gg A/\omega_1$. The finite interaction is also responsible for the appearance of two peculiar points at $\omega=\omega_2\pm A/(2\omega_2)$. At the point $\omega=\omega_2- A/(2\omega_2)$ the amplitude reaches its maximum. Oppositely,  the oscillation amplitude goes to zero at the frequency $\omega=\omega_2+ A/(2\omega_2)$.  Such a reduction of the oscillation amplitude is called the dynamical damping and is very well known in the oscillation theory. Two periodic forces act on the the magnetic moment $\mathbf m_1$. The first one is due to the external field and the second one is due to the interaction with the second magnetic layer. Phases of the forces depend on frequency. When the phase difference is $\pi$ the forces cancel each other. Such a cancellation appears at $\omega=\omega_2+ A/(2\omega_2)$ and therefore, $\mathbf m_1$ does not oscillate at this frequency. At $\omega=\omega_2-A/(2\omega_2)$ these two  forces are in phase leading to enhancement of oscillations. Finally, the shape of the resonance peak is distorted and the peak asymmetry appears. Such a peculiarity in the frequency dependence of the oscillation amplitude is well known as the Fano resonance~[\onlinecite{Satanin2006}].  

\begin{figure}
	\includegraphics[width=0.5\columnwidth]{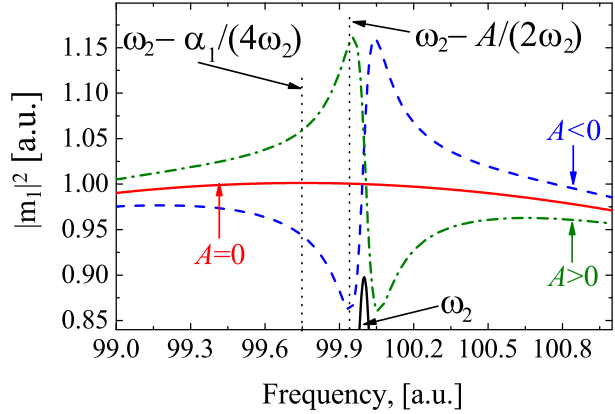}
	\caption{Amplitude of the magnetization of the first layer $|m_1|^2$ as a function of frequency $\omega$. The red line is for the zero interlayer coupling ($J=0$). The blue dashed line is for the finite AFM interaction ($J<0$). The green dash-dotted line corresponds to $J>0$.  The black line shows the amplitude of the second layer oscillation $|m_2|^2$ (reduced 10 times to make it comparable to $|m_1|^2$).\label{Fig:Fano1}}%
\end{figure}

When we take finite $\alpha_2$ into account there is no full damping and the amplitude is not zero, but one still has the minimum at $\omega=\omega_2+A/(2\omega_2)$ and the maximum at $\omega=\omega_2-A/(2\omega_2)$. 

Important feature here is that if one changes the interaction sign the minimum and maximum switch their positions. For $A<0$ (antiferromagnetic (AFM) interaction) the dynamical damping appears below $\omega_2$. For $A>0$ (FM interaction) the dynamical damping appears above $\omega_2$. This feature can be used for defining the interaction sign.

Figure~\ref{Fig:Fano1} demonstrates behavior of $|m_1|^2$ as a function of frequency for $\omega_2=100$ a.u. and $\alpha_1=10$ a.u. The solid red curve shows the case of zero interaction, $A=0$. In this case there are no peculiarities in the amplitude behavior. Blue dashed curve in Fig.~\ref{Fig:Fano1} shows $|m_1|^2$ for finite AFM interaction $A=-3$ a.u. One can easily see the asymmetry of the resonant peak. According to our consideration the dynamical damping occurs in this case below $\omega_2=100$ a.u.  Note that the curve is plotted for finite $\alpha_2$ and therefore instead of zero amplitude at $\omega=\omega_2- A/(2\omega_2)$ we have finite oscillations. The dynamical enhancement appears at $\omega=\omega_2+ A/(2\omega_2)$. Dash-dotted green line shows $|m_1|^2$ for positive FM interaction $A=3$ a.u. One can see that the Fano resonance (asymmetry) is reflected with respect to $\omega=\omega_2$ in this case. So, the shape of the peak is clearly different for different sign of the interlayer interaction.

Closing this section we have to mention that the Fano resonance disappears if the dissipation is the same in both layers. 

\begin{figure}
	\includegraphics[width=0.5\columnwidth]{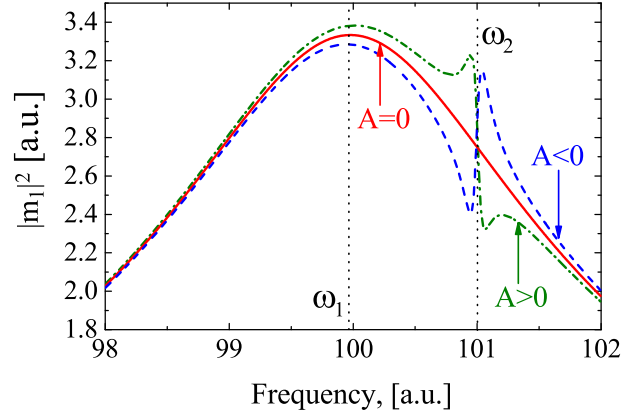}
	\caption{Amplitude of the magnetization of the first layer $|m_1|^2$ as a function of frequency $\omega$. The case when the resonant frequencies of the layers are different. Red line is for the zero interlayer coupling ($J=0$). Blue dashed line is for the finite AFM interaction ($J<0$). Green dash-dotted line corresponds to the finite FM interaction ($J>0$).\label{Fig:Fano2}}%
\end{figure}

\subsection{Layers with essentially different resonant frequencies} 
Similar behavior occurs when  the resonant frequencies of two layers are not the same.  The Fano resonance appears around the resonant frequency of the layer with lower dissipation. Again, the sign of the interlayer interaction defines the shape (``direction'') of the Fano resonance. Fig.~\ref{Fig:Fano2} shows the amplitude $|m_1|^2$ as a function of frequency for $\omega_1=100$ a.u. and $\omega_2=101$  a.u., $\alpha_1=3$  a.u. and $\alpha_2=0.1$  a.u., $A=0,\pm 3$  a.u.

Important to note that the Fano  peculiarity disappears as the resonance frequencies become far from each other and there is no overlap between the FMR peaks.

\subsection{Absorption.}
In the FMR experiment the measured quantity $W$ is the absorption or imaginary part of the system response 
\begin{equation}\label{Eq:Abs}
W/\omega=M_0h^{(0)}\mathrm{Im}(m_{1x}+m_{2x}) \sim \alpha_1|m_{1x}|^2+\alpha_2|m_{2x}|^2.
\end{equation}

\begin{figure}
	\includegraphics[width=0.5\columnwidth]{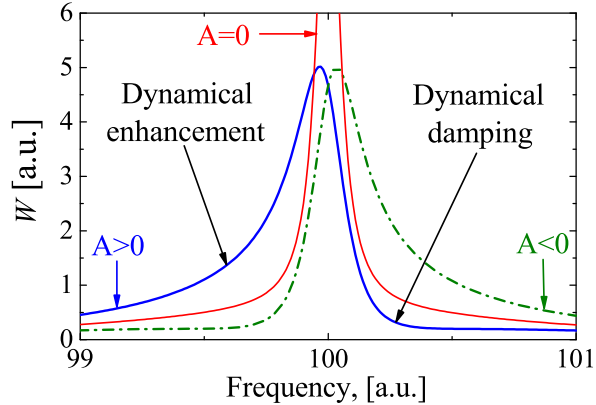}
	\caption{Absorption $W$ (Eq.~(\ref{Eq:Abs})) as a function of frequency $\omega$. The case of equal resonant frequencies of the magnetic layers is shown. Red line is for the zero interlayer interaction ($J=0$). Blue solid line is for finite FM interaction ($J>0$). Green dash-dotted line corresponds to finite AFM interaction ($J<0$).\label{Fig:Abs}}%
\end{figure}

Figure~\ref{Fig:Abs} shows the absorption as a function of frequency for two interacting magnetic moments. Resonance frequencies are $\omega_{1,2}=100$ a.u., $\alpha_1 = 0.1$ a.u., $\alpha_2=2$ a.u., $A=0,\pm 50$ a.u. One can see that at zero interaction the absorption peak is symmetric, while for finite interaction the peak asymmetry appears. At that the asymmetry is defined by the interlayer interaction sign. 

\section{Numerical simulations}\label{Sec:NumSim}
In the previous section on the base of the simplified model it was shown that the FMR peak asymmetry arises due to a weak interaction of the magnetic layers. The frequency dependencies of FMR signal were studied which was relevant for comparison of magnetic multilayer systems with other systems showing the Fano resonances. In the FMR experiment the field dependence is ordinarily measured at a fixed frequency of alternating field.

Besides, in the model a limit of strong field was considered in which magnetizations $\mathbf M_{1,2}$ were co-directed with each other and with the external field. In a real FMR experiment the magnitude of the external field is limited. Therefore, the coincidence of resonance fields of the magnetic layers ($H_{\mathrm r1}\approx H_{\mathrm r2}$) may appear in the situation when the external magnetic field and the equilibrium magnetic moments of the layers are not co-directed. Analytical solution of the problem in this situation is not feasible. Therefore, here we present numerical demonstration of the FMR peak asymmetry in a realistic situation. 

\begin{figure}
	\includegraphics[width=0.5\columnwidth]{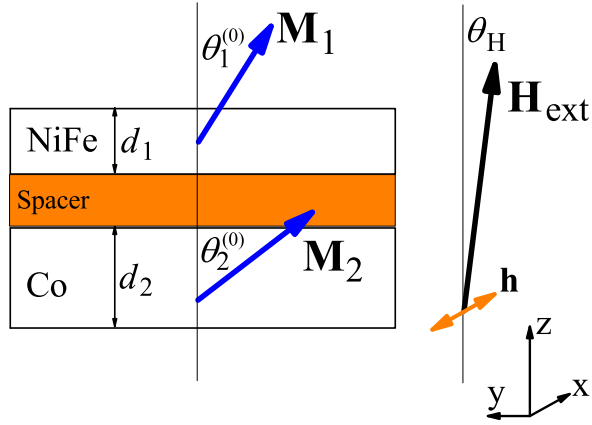}
	\caption{System geometry used in our numerical modeling. Two magnetic layers (NiFe and Co) with thicknesses $d_{1,2}$ are placed in an external magnetic field $\mathbf H_{\mathrm{ext}}$. The field makes angle $\theta_\mathrm H$ with the layers normal. The alternating magnetic field $\mathbf h$ is applied perpendicular to $\mathbf H_\mathrm{ext}$.  Equilibrium magnetic moments $\mathbf M_{1,2}$ make angles $\theta^{(0)}_{1,2}$ with the normal. \label{Fig:ExpGeom}}%
\end{figure}

\begin{figure}
	\includegraphics[width=0.5\columnwidth]{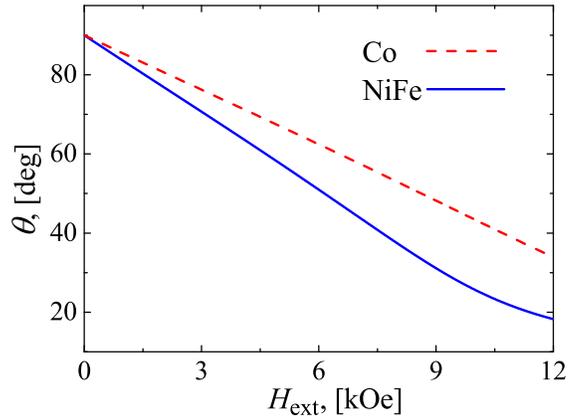}
	\caption{Equilibrium angles $\theta^{(0)}$ for Co and NiFe layers as a function of external field magnitude. The external field is applied by the angle $\theta_H=5.8$ deg with respect to the sample normal. \label{Fig:EquilAngles}}%
\end{figure}

We use a well known numerical algorithm to solve the LLG equations for magnetic films~[\onlinecite{Baberschke2003,Ozdemir2010}]. The system energy is given by

\begin{equation}\label{Eq:SysEn}
E=E_\mathrm Z+E_\mathrm D+E_{\mathrm A}+E_{\mathrm{int}},
\end{equation}
where the Zeeman energy is 
\begin{equation}\label{Eq:EnZee}
E_\mathrm Z=-\sum_{i=1,2}d_i(\mathbf M_i\mathbf H_\mathrm{ext}),
\end{equation}
the magneto-dipole shape anisotropy is
\begin{equation}\label{Eq:EnShape}
E_\mathrm D=\sum_{i=1,2}2\pi d_i M_i^2\cos^2(\theta_i),
\end{equation}
the uniaxial anisotropy is
\begin{equation}\label{Eq:EnAn}
E_\mathrm A=\sum_{i=1,2}d_i K_i\cos^2(\theta_i).
\end{equation}
Here $\theta_{1,2}$ are the polar angles of magnetizations (see Fig.~\ref{Fig:ExpGeom}). The external magnetic field $\mathbf H_\mathrm{ext}$ is inclined by an angle $\theta_\mathrm H$ with respect to the sample normal. $K$ is the anisotropy constant. Equilibrium angles of magnetizations $\theta^{(0)}_{1,2}$ are defined by minimization of the system energy Eq.~(\ref{Eq:SysEn}). We use the parameters approximately corresponding to the NiFe/I/Co magnetic bilayer. The thickness of NiFe and Co is $d=1$ nm, g-factors are $g_{1,2}=2$, the frequency of the alternating field is $\omega=9.5$ GHz, the saturation magnetizations are $M_1=325$ Gs, $M_2=1420$ Gs, the uniaxial anisotropy constants are $K_1=-7.5\cdot 10^5$ Gs$\cdot$Oe and $K_2=4\cdot 10^6$ Gs$\cdot$Oe, the damping parameters are $\alpha_1=0.006$ and $\alpha_2=0.04$.

Figure \ref{Fig:EquilAngles} shows behaviour of equilibrium magnetization angles as a function of the external field magnitude at $\theta_H=5.8$ deg. The field magnitude and angle are chosen in the region where we will observe the FMR peak asymmetry. One can easily see that the equilibrium magnetic moments are not co-directed with each other and with the magnetic field.

\begin{figure}
	\includegraphics[width=0.5\columnwidth]{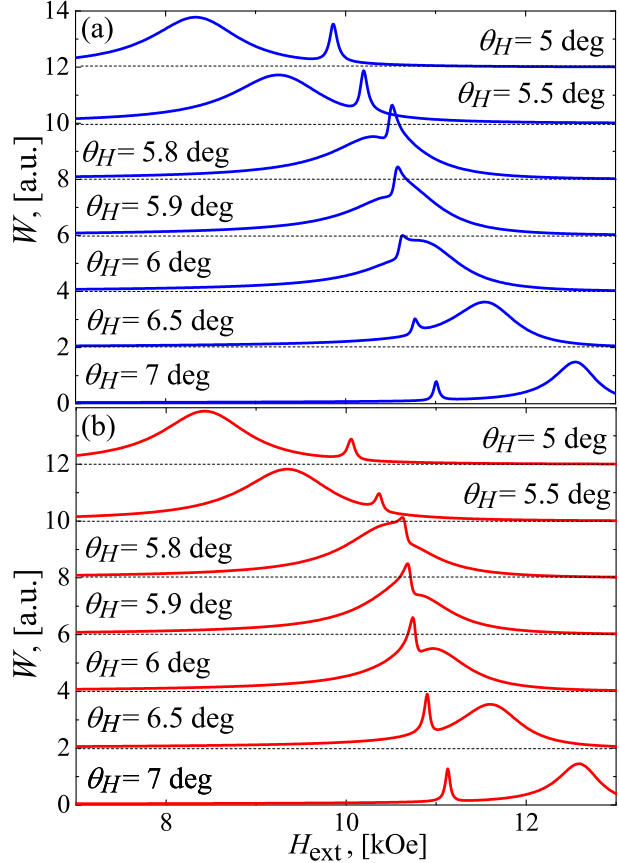}
	\caption{FMR spectrum (absorbed power $W$ as a function of the external field magnitude $H_\mathrm{ext}$) obtained numerically for NiFe/Co system. (a) FM interlayer interaction $\tilde J=0.001$ J$/$m$^2$. (b) AFM interlayer interaction $\tilde J=-0.001$ J$/$m$^2$. Differnet curves in the same plot correspond to different inclination angle of the external magnetic field $\theta_H$. The curves for different $\theta_H$ are shifted with respect to each other for better visibility.  \label{Fig:FMRnum}}%
\end{figure}

Figure \ref{Fig:FMRnum} shows the dependence of the FRM signal as a function of the external magnetic field magnitude ($W(H_\mathrm{ext})$) at a fixed frequency of the alternating field. The upper and lower panels correspond to different sign of the exchange interaction $\tilde J=\pm 0.001$ J$/$m$^2$. Each figure shows several plots for different angle $\theta_H$ of the applied field. When the angle $\theta_H>6.5$ deg and $\theta_H<5.5$ deg, one sees two separate peaks corresponding to NiFe and Co layers. The NiFe peak is the narrow one and the Co peak is the wide one. Changing the angle of the applied field one shifts the resonance field of NiFe and Co films $H_{\mathrm r1,2}$. Since the magnetic anisotropy of these films is quite different $H_{\mathrm r1,2}(\theta_H)$ the dependencies are not the same and intersect each other at a certain angle $\theta_H$. One can see that peaks overlap at the angle $\theta_H\approx 5.9$ deg.

There is no peak asymmetry when NiFe and Co peaks are far from each other. This is in agreement with our analytical model. The asymmetry appears when the peaks overlap. Comparing upper an lower panel one can see that the peak asymmetry is different for FM and AFM interaction. Therefore, one can define the interaction sign by measuring FMR spectrum at conditions of intersection of peaks. If the slope of the narrow peak is higher on the left side the interaction is FM. If the slope is higher on the right side the interaction is AFM. Fitting experimental data one can even define the magnitude of the interlayer interaction.

\section{Conclusion}
We considered the FMR resonance in two coupled magnetic layers. We showed that the interaction between these layers leads to the occurrence of the so-called Fano resonance. The Fano resonance shows as a peculiarity in the absorption spectrum of the coupled system. In particular, the resonance peak becomes asymmetric. The asymmetry type is defined by the sign of the interaction between the layers. One can use the asymmetry to distinguish between FM and AFM interlayer coupling. Using numerical simulations one can even estimate a magnitude of the interaction fitting the asymmetric FMR peak.

As a final remark we would like to mention that in our work we considered the isotropic interaction Eq.~(\ref{Eq:InterGen}). Such an equation describes the exchange coupling. However, many experiments evidence that in magnetic multilayer systems there is also the magneto-dipole coupling called the ``orange-peel'' effect. In contrast to the exchange coupling, the ``orange-peel'' effect is anisotropic and described by a different equation~[\onlinecite{Fraerman2018}]. The anisotropy will lead to the angular dependence of the coupling constant $J=J(\theta_H)$. This, peculiarity can be used for distinguishing between the exchange coupling and the ``orange-peel'' effect. This opportunity requires further investigation. 

\section{Acknowledgments}
This research was supported was supported by the Russian Science Foundation (Grant  16-12-10340).

\bibliography{FMR}

\end{document}